\documentclass{ws-procs11x85}
\usepackage{ws-procs-thm}           

\begin{document}

\title{Do AI chatbots find what experts would? Effects of model, user role, and sample size on study retrieval for medical questions}

\author{
Qingfang Liu$^{1}$,
Qiao Jin$^{2}$,
Joe D. Menke$^{3}$,
Thorsten Kahnt$^{1}$,
Zhiyong Lu$^{2,\dagger}$
}

\address{
$^{1}$National Institute on Drug Abuse Intramural Research Program,
National Institutes of Health, Baltimore, MD, USA\\
$^{2}$National Library of Medicine,
National Institutes of Health, Bethesda, MD, USA\\
$^{3}$School of Information Sciences,
University of Illinois Urbana--Champaign, Champaign, IL, USA\\
$^{\dagger}$Corresponding author: zhiyong.lu@nih.gov
}

\copyrightinfo{\copyright\ 2026 The Authors. Open Access chapter published by World Scientific Publishing Company and distributed under the terms of the Creative Commons Attribution Non-Commercial (CC BY-NC) 4.0 License.}

\vspace{0.5em}

\begin{abstract}
Large language model (LLM) chatbots are increasingly used to answer clinical questions with citations to relevant clinical studies. Prior research has largely focused on citation fabrication, leaving a gap in evaluating the quality of the retrieved studies and the factors driving their selection, particularly for newer models with stronger reasoning capabilities. In this study, we evaluated three recent, general-purpose LLM chatbots: Claude Sonnet 5, Gemini 3.1 Pro, and ChatGPT GPT-5.5. We prompted the models with clinical questions adapted from 20 review questions in Issues 6 and 7 of the 2026 Cochrane Database of Systematic Reviews, simulating patient, clinician, and evidence-synthesis researcher user roles. For each review question, we queried each of the three chatbots under each of the three user roles, with four independent repetitions per chatbot--role combination, yielding 720 responses in total ($3$ chatbots $\times$ $3$ user roles $\times$ $4$ repetitions $\times$ $20$ review questions). Each chatbot was asked to support its answers with primary clinical citations, which we then benchmarked against the included and excluded study sets of the corresponding Cochrane reviews. Included studies were those judged by Cochrane reviewers to meet the review-specific eligibility criteria, whereas excluded studies did not meet one or more of these criteria but could still be relevant to the clinical question posed to the chatbot. On average, a single chatbot response retrieved 39.2\% ± 29.8\% (mean $\pm$ SD across all 720 responses) of the corresponding Cochrane included-study set and 5.0\% $\pm$ 9.4\% of the corresponding Cochrane excluded-study set. Recall of Cochrane included studies varied significantly by model and user role. ChatGPT achieved higher recall than Claude or Gemini (63.1\% $\pm$ 29.5\% vs.\ 37.0\% $\pm$ 23.8\% vs.\ 17.3\% $\pm$ 13.1\%; blocked permutation test, $p=2.0\times10^{-5}$). The researcher role yielded higher recall than the clinician or patient roles (42.8\% $\pm$ 30.8\% vs.\ 38.6\% $\pm$ 28.9\% vs.\ 36.1\% $\pm$ 29.3\%; $p=2.0\times10^{-5}$). Controlling for publication year, citations per year, and open-access status, sample size was the only significant predictor of retrieval, with each doubling of sample size associated with 50\% higher odds of retrieval (odds ratio 1.50, 95\% CI 1.24--1.81, $p=2.34\times10^{-5}$). These findings suggest that while LLM chatbots can retrieve some studies identified by expert reviewers, their performance varies by model and user role, and they exhibit a bias toward clinical trials with larger sample sizes. Collected LLM responses, analysis code, and supplementary materials are available at https://github.com/QingfangLiu/llm-evidence-retrieval-bias.
\end{abstract}


\clearpage

\section{Introduction}\label{aba:sec1}

An increasing number of individuals are consulting chatbots to address medical questions, including requests to retrieve evidence from existing literature
to better understand medical conditions \cite{costagomes2026healthqueries,
shpiner2026chatbotfirst}, encouraged in part by strong chatbot performance
on medical knowledge benchmarks \cite{singhal2023clinicalknowledge, Jin2026LLMMedicalResearch}. Yet
the ability of general-purpose chatbots to identify and cite relevant
clinical studies remains less well characterized \cite{somer2026studyidentification,
chelli2024hallucination, gwon2024scientificsearches,sarker2024natural}, especially in light
of prior findings that large language models can fabricate or hallucinate
citations \cite{walters2023fabrication, wu2025sourcecheckup, wang-etal-2025-medcite, jin2026med, topaz2026fabricated}. At the same
time, systematic reviews and meta-analyses synthesize primary studies
selected by experts to produce higher-quality conclusions for medical
questions \cite{lefebvre2025searching}. This raises an important question:
how closely do the studies cited by chatbots align with those included in
systematic reviews?

Some prior work has begun to evaluate how well general-purpose chatbots retrieve primary studies, generally by comparing chatbot-cited studies
against the included-study lists of published systematic
reviews. These evaluations generally report low recall, whereas precision and citation fabrication vary across systems and evaluation methods. Somer et al.
\cite{somer2026studyidentification} evaluated six publicly accessible systems against the 14 studies in an obstetric meta-analysis; the best-performing system, Claude 3.7, identified five studies. Chelli et al.
\cite{chelli2024hallucination} tested three LLMs across 11 systematic reviews of rotator-cuff interventions, reporting precision of 0--13.4\% and hallucination rates of 28.6--91.4\%. Gwon et al.
\cite{gwon2024scientificsearches} found that ChatGPT and Bing AI identified
only one and two benchmark randomized trials, respectively, out of 24 randomized trials. Sidhu et al.
\cite{sidhu2026trust} compared the authenticity, quality, and geographic
provenance of over 1{,}200 references across nine contemporary chatbot
configurations. Low et al. \cite{low2025clinicalquestions} found that general-purpose LLMs produced very few relevant, evidence-based answers (2--10\% of questions). However, many of these studies relied on models predating advanced reasoning and agentic systems, leaving it unclear how modern, more capable models behave in these contexts. Additionally, much prior work has focused on a single medical domain \cite{somer2026studyidentification, chelli2024hallucination, gwon2024scientificsearches}, limiting the generalizability of their conclusions.

Outside biomedicine, retrieval-benchmark work such as LitSearch \cite{ajith2024litsearch} showed that retrieval performance varies substantially with query specificity and query construction. PaperAsk evaluated GPT-4o, GPT-5, and Gemini-2.5-Flash across a range of scientific fields and found that citation retrieval failed in 48-98\% of multi-reference queries \cite{wu2025paperaskbenchmarkreliabilityevaluation}. Gao et al. evaluated five LLMs for 40 randomly selected original articles and found that the models failed to retrieve correct reference data 47.8\% of the time \cite{gao2026errors}. Related work has probed adjacent failure modes rather than recall itself: whether generated citations actually support their associated claims \cite{wu2025sourcecheckup}, how often citations are outright fabricated or contain metadata errors \cite{walters2023fabrication}, and how reference-generation performance varies by model, discipline, and publication recency across large literature-review corpora
\cite{tang2025literaturereview}. Other work asked chatbots to generate search strategies for bibliographic databases such as PubMed and Embase rather than to select studies directly, but this setup may not reflect how general users interact with chatbots \cite{tam2026database}.

A further consideration is that different types of users (e.g., patients, clinicians, evidence synthesis researchers) may seek access to primary clinical studies for medical question answering \cite{roberts2016interactive,li2014reasons}, and whether such variation affects evidence retrieval remains an open question. Task-aligned role-play has been shown to help with reasoning benchmarks \cite{kong2024roleplay}, while assigning the model a persona does not reliably improve accuracy and can sometimes reduce it \cite{zheng2024personas}. Persona information more broadly shifts model predictions in ways that do not always track genuine human variation \cite{hu2024persona}. Prompt architecture and framing can induce systematic, non-neutral bias even when no persona is involved \cite{brucks2025promptarchitecture}, and patient question framing alone has been shown to change a model's medical conclusions even when the underlying evidence is held fixed \cite{yun2026framing}. However, no prior study has tested whether a user's self-identified role changes which primary studies a chatbot retrieves.

In this study, we evaluated three contemporary general-purpose chatbot systems (GPT-5.5, Claude Sonnet 5, and Gemini 3.1 Pro, all equipped with the ability to perform web searches) on medical questions adapted from 20 Cochrane review topics. For each topic, prompts were framed from the perspective of a patient, clinician, or
evidence-synthesis researcher, and each role--topic combination was repeated
four times. GPT-5.5 achieved the highest recall of Cochrane included-study sets, and
evidence-synthesis researcher framing yielded higher recall than clinician or
patient framing. After controlling for publication year, citation rate, and open-access status, larger sample size remained the only significant predictor of study retrieval. These findings indicate that chatbot-based evidence retrieval is incomplete, context-dependent, and biased toward larger clinical trials, with implications for patients who use chatbots for medical information, as well as for clinicians, LLM developers, medical AI researchers, and policymakers \cite{meyer2023chatgpt,weissenbacher2026enhancing,sahoo2024large}.

\section{Methods}
\subsection{Review selection}

We used Cochrane intervention reviews published in Issues 6 and 7 of the 2026 \textit{Cochrane Database of Systematic Reviews}. These were the two most recent complete issues available at the time of the experiment and were published after the knowledge cutoff dates of all three chatbots (see below). The 20 eligible intervention reviews covered diverse clinical areas, evaluating pharmacological or biologic interventions (n = 7), procedural, surgical, or laboratory techniques (n = 4), bedside-care or clinical-management strategies (n = 3), nutritional supplementation (n = 2), rehabilitative or conservative physical treatments (n = 2), behavioral interventions (n = 1), and telehealth or service-delivery interventions (n = 1). We excluded protocols and reviews addressing diagnostic, prognostic, or other non-intervention questions. 

\subsection{Experiment}

We evaluated three widely used consumer chatbots using the
reasoning-enabled model configurations in
\textbf{Table~\ref{tab:chatbot-configurations}}. We focused on the general-purpose chat interfaces rather than specialized Deep Research modes to evaluate evidence retrieval in ordinary question-answering settings rather than extended literature search. To characterize the broader capability of the evaluated model configurations, we also report performance on Humanity's Last Exam (HLE), a multimodal, frontier-level academic benchmark comprising 3,000 expert-authored questions across dozens of subjects and designed to probe advanced knowledge and reasoning. These scores show that the evaluated systems are broadly comparable frontier-level models, though HLE itself does not benchmark literature retrieval, the primary focus of the current work. Each response was generated independently in a fresh anonymous web session with no prior conversational context. Data were collected from the U.S. East Coast between mid- and late July 2026.

\begin{table}[h]
\tbl{Chatbot systems, configurations, and benchmark performance.}
{\small
\begin{tabular}{@{}
p{0.20\textwidth}
p{0.13\textwidth}
p{0.14\textwidth}
p{0.20\textwidth}
p{0.10\textwidth}
p{0.10\textwidth}
@{}}
\toprule
\raggedright Model &
\raggedright Knowledge cutoff &
\raggedright Release date &
\raggedright Reasoning configuration &
\raggedright HLE (no tools) &
\raggedright HLE (with tools)\tabularnewline
\colrule
\raggedright Anthropic Claude Sonnet 5 &
\raggedright January 2026 &
\raggedright June 30, 2026 &
\raggedright Medium effort; thinking enabled &
\raggedright 43.2\% &
\raggedright 57.4\%\tabularnewline

\raggedright Google Gemini 3.1 Pro &
\raggedright January 31, 2025 &
\raggedright February 19, 2026 &
\raggedright Extended thinking &
\raggedright 44.4\% &
\raggedright 51.4\%\tabularnewline

\raggedright OpenAI GPT-5.5 &
\raggedright December 1, 2025 &
\raggedright April 23, 2026 &
\raggedright High reasoning &
\raggedright 41.4\% &
\raggedright 52.2\%\tabularnewline
\botrule
\end{tabular}}
\tabnote{Note: Knowledge cutoff indicates the latest date through which a model's documented training knowledge extends. HLE (Humanity's Last Exam) is a multimodal benchmark of advanced knowledge and reasoning and is reported here to provide external context on general model capability; it is not a benchmark of evidence retrieval.}
\label{tab:chatbot-configurations}
\end{table}

For each review question, we developed three prompts representing different user roles. Except for the user role stated at the beginning of each prompt, the wording was kept as consistent as possible. We also explicitly instructed the chatbots to rely on primary evidence (e.g., original research studies and clinical trials) rather than secondary evidence (e.g., systematic reviews, meta analyses, narrative reviews, committee opinions, practice guidelines, editorials, or commentaries) when answering. Given that the chatbot systems had web-search access, this instruction was intended to avoid, or where that was not possible, reduce reliance on secondary sources, including the benchmark Cochrane reviews, when identifying studies to present in the response. To account for variability in chatbot responses, each prompt was run four times. An example is provided in \textbf{Table~\ref{tab:prompt}}.

\begin{table}[ht]
\tbl{Example of the three user-role prompt variants for one review question.}
{\small
\begin{tabular}{p{0.22\textwidth} p{0.70\textwidth}}
\toprule
User role & Role-specific opening \\
\colrule
Patient &
I have high blood pressure, and I am considering taking medicine to help me lose weight. \\
Clinician &
I am a clinician caring for people with high blood pressure who are considering taking medicine to help them lose weight. \\
Evidence-synthesis researcher &
I am an evidence-synthesis researcher studying weight-loss medicines in people with high blood pressure. \\
\colrule
\multicolumn{2}{p{0.94\textwidth}}{Shared remainder of the prompt: I am trying to understand whether weight-loss medicines improve the health of people with high blood pressure. Find individual primary studies on this. Do not use systematic reviews, meta-analyses, narrative reviews, committee opinions, practice guidelines, or editorials or comments. List the primary studies at the end.} \\
\botrule
\end{tabular}}
\label{tab:prompt}
\end{table}

\subsection{Study identification}

We used the full set of studies listed as ``included" and ``excluded" in the corresponding Cochrane review as the reference standard and analyzed the studies retrieved in each chatbot response. For each response, the unit of analysis was the set of uniquely identifiable primary studies referenced, rather than the number of citations or publications, consistent with the distinction between studies and publications in standard Cochrane methodology. 

Study identification was performed in two stages. First, we used an AI-assisted workflow to extract citation information and propose candidate matches. Second, one author manually verified all proposed matches against the Cochrane review's study records and, where necessary, external bibliographic databases. Because the prompts did not specify a citation format, we considered a study identifiable when the information provided (e.g., a study or trial name, author and year, title, PMID or DOI, registry identifier, or the destination of a citation link) was sufficient to resolve the underlying study. A citation with minor metadata errors (such as a wrong volume number) was still counted as a match, provided the rest of the information clearly pointed to one specific study; we report the metadata errors below. Unresolved citations received no matched-study credit. 

For each chatbot response, we also assessed compliance with the instruction to rely on primary evidence by recording whether the response referenced or discussed a Cochrane review. For each Cochrane-included study, we resolved PMIDs for linked reports
using RIS metadata and PubMed searches, then obtained citation counts
from Semantic Scholar. We used the highest available citation count
among linked reports to calculate citations per year, dividing by years
since the Cochrane-reported publication year (minimum one year).
Open-access status was the Semantic Scholar indicator for the report
providing that citation count.

\section{Results}
\subsection{Number and composition of retrieved studies}

\begin{figure}[ht]
\centerline{\includegraphics[width=0.95\textwidth]{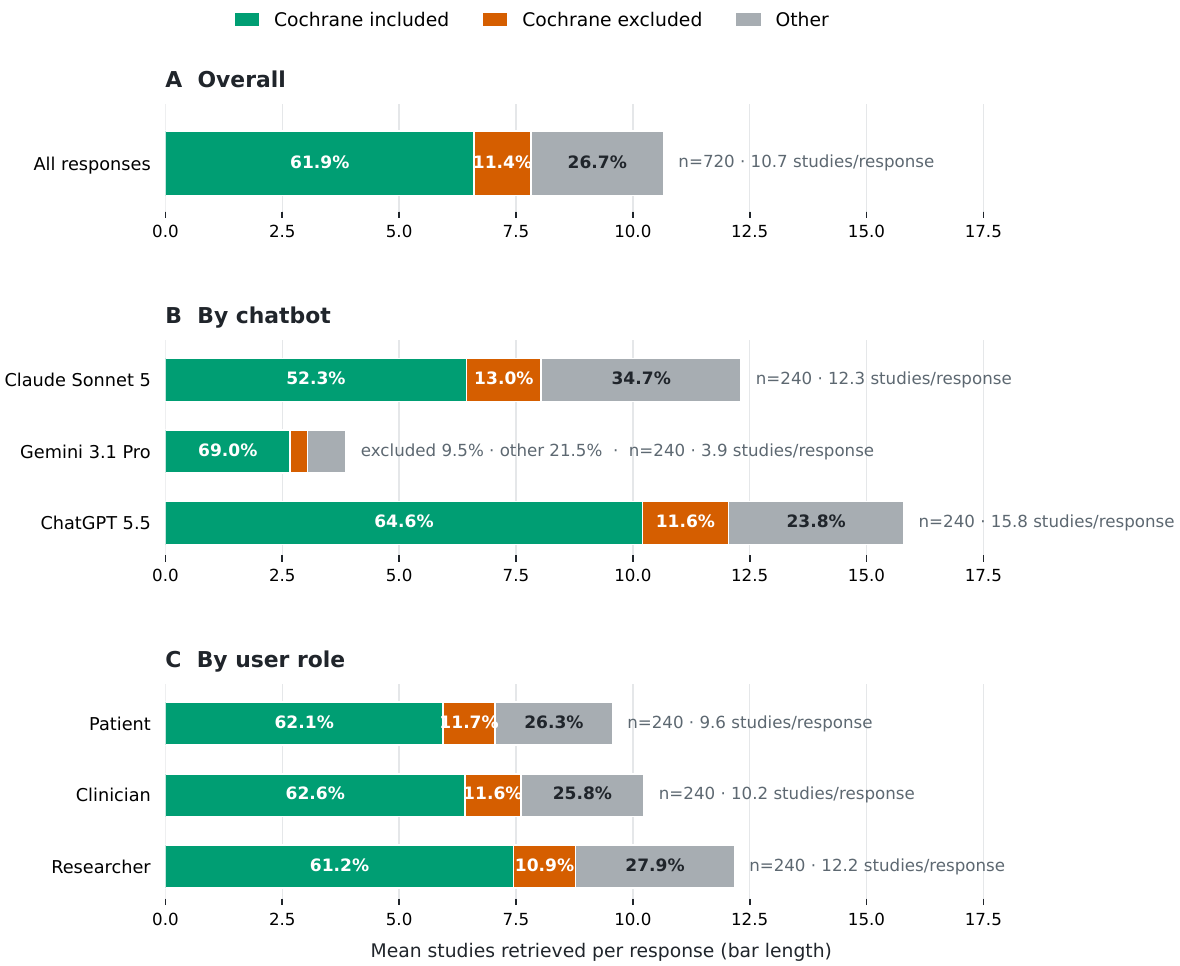}}
\caption{\textbf{Composition and volume of studies retrieved by Cochrane status.}
Stacked bars show the mean within-response proportion of distinct retrieved
studies classified as Cochrane included, Cochrane excluded, or other, with
bar length encoding mean studies retrieved per response on one shared scale
across panels. Panel A summarizes all 720 responses; Panels B and C stratify
by chatbot and user role, respectively (240 responses per group).}
\label{fig:candidate-status-composition}
\end{figure}

Across 720 responses, the mean number of studies retrieved per response was 10.66. Of the studies retrieved in each response, a mean of 61.9\% were Cochrane-included studies, compared with 11.4\% Cochrane-excluded and 26.7\% classified as other. Both retrieval volume and study composition varied across chatbots (\textbf{Figure~\ref{fig:candidate-status-composition}}). Claude retrieved a mean of $12.32 \pm 6.24$ studies per response, of which 52.3\% were classified as included---the lowest included-study proportion among the three chatbots. Gemini retrieved the fewest studies per response ($3.87 \pm 1.66$) but had the highest included-study proportion (69.0\%). ChatGPT retrieved the most studies per response ($15.80 \pm 6.60$), of which 64.6\% were classified as included. In contrast to the chatbot-level differences, study composition was similar across user roles, although researcher-role responses retrieved more studies than patient- or clinician-role responses.

\begin{figure}[ht]
\centerline{\includegraphics[width=\textwidth]{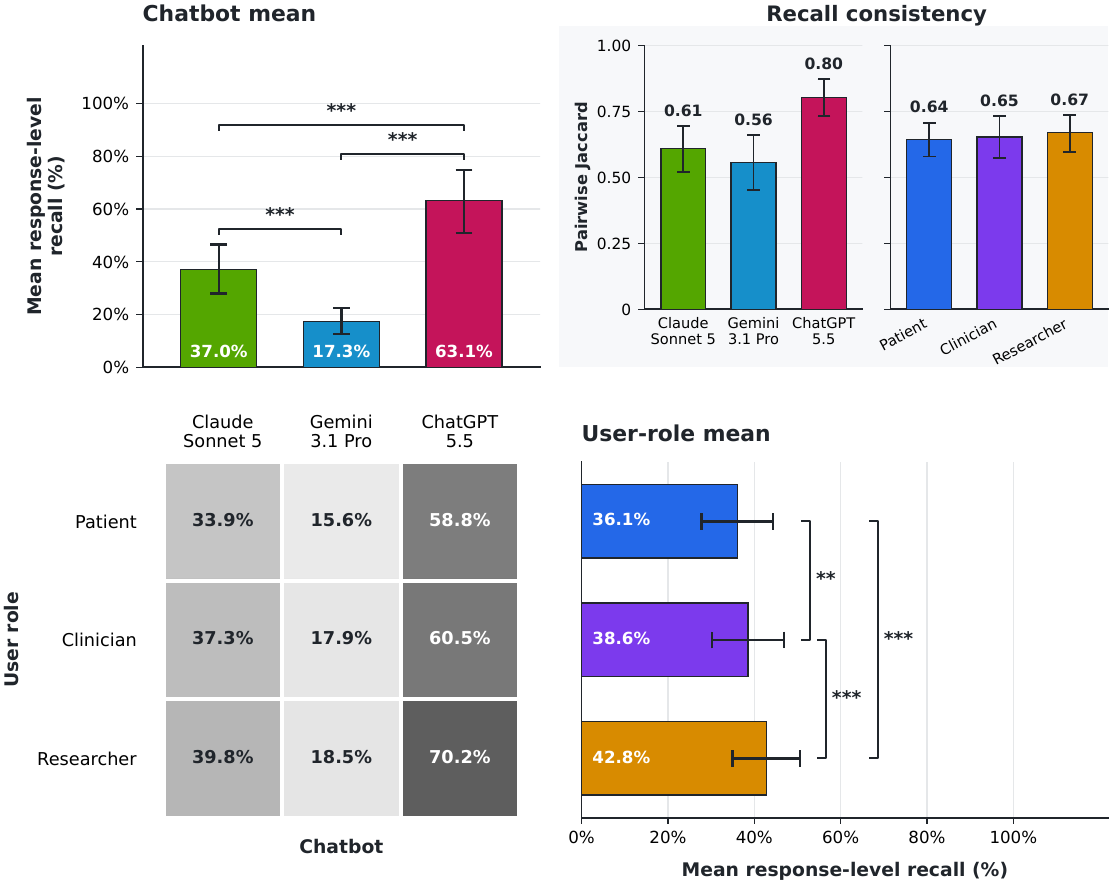}}
\caption{\textbf{Cochrane included-study recall and recall consistency by chatbot and user role.}
The heatmap shows mean response-level recall for each chatbot-by-user-role combination; marginal bars summarize recall by chatbot and by user role. The top-right panel shows recall consistency, defined as the mean pairwise Jaccard similarity of Cochrane-included study sets across replicate responses. Error bars indicate percentile 95\% review-clustered bootstrap confidence intervals. Recall was calculated as the proportion of Cochrane-included studies retrieved in each response. Pairwise differences were tested using blocked permutation tests (50,000 permutations per comparison), with $p$-values adjusted by the Holm method within each marginal dimension. Pairwise differences in mean recall (percentage points; 95\% review-clustered bootstrap CIs from 50,000 resamples) were: Claude minus Gemini, 19.7 (14.3--25.7); ChatGPT minus Claude, 26.1 (19.3--33.5); ChatGPT minus Gemini, 45.8 (36.3--55.0); clinician minus patient, 2.5 (0.9--4.2); researcher minus clinician, 4.3 (2.5--6.2); and researcher minus patient, 6.8 (4.6--9.1). ${}^{***}P<0.001$; ${}^{**}P<0.01$; ${}^{*}P<0.05$; ns, $P\geq0.05$.}
\label{fig:retrieval-recall}
\end{figure}

\subsection{Included-study recall and recall consistency}

Across 720 responses, the mean per-response recall of Cochrane-included studies was $39.2\% \pm 29.8\%$ (mean $\pm$ SD), while the mean proportion of Cochrane-excluded studies cited was $5.0\% \pm 9.4\%$. Differences in mean included-study recall by chatbot and user role were assessed using separate blocked permutation tests with 50,000 permutations each. Mean included-study recall differed significantly by chatbot: $37.0\% \pm 23.8\%$ for Claude Sonnet 5, $17.3\% \pm 13.1\%$ for Gemini 3.1 Pro, and $63.1\% \pm 29.5\%$ for ChatGPT 5.5 ($p=2.0 \times 10^{-5}$). Recall also differed by user role: $36.1\% \pm 29.3\%$ for patient, $38.6\% \pm 28.9\%$ for clinician, and $42.8\% \pm 30.8\%$ for researcher ($p=2.0 \times 10^{-5}$). Recall consistency, measured as mean pairwise Jaccard similarity of retrieved Cochrane-included study sets across replicate responses, was highest for ChatGPT 5.5 (0.80), compared with Claude Sonnet 5 (0.61) and Gemini 3.1 Pro (0.56), and was similar across user roles (patient, 0.64; clinician, 0.65; researcher, 0.67). These results are summarized in \textbf{Figure~\ref{fig:retrieval-recall}}. Although our primary focus was on recall-based metrics, we also report a comparison of Cochrane-included-study precision in the Supplementary (\textbf{Figure S1}).

\subsection{Study overlap by chatbot and user role}

\begin{figure}[ht]
\centerline{\includegraphics[width=\textwidth]{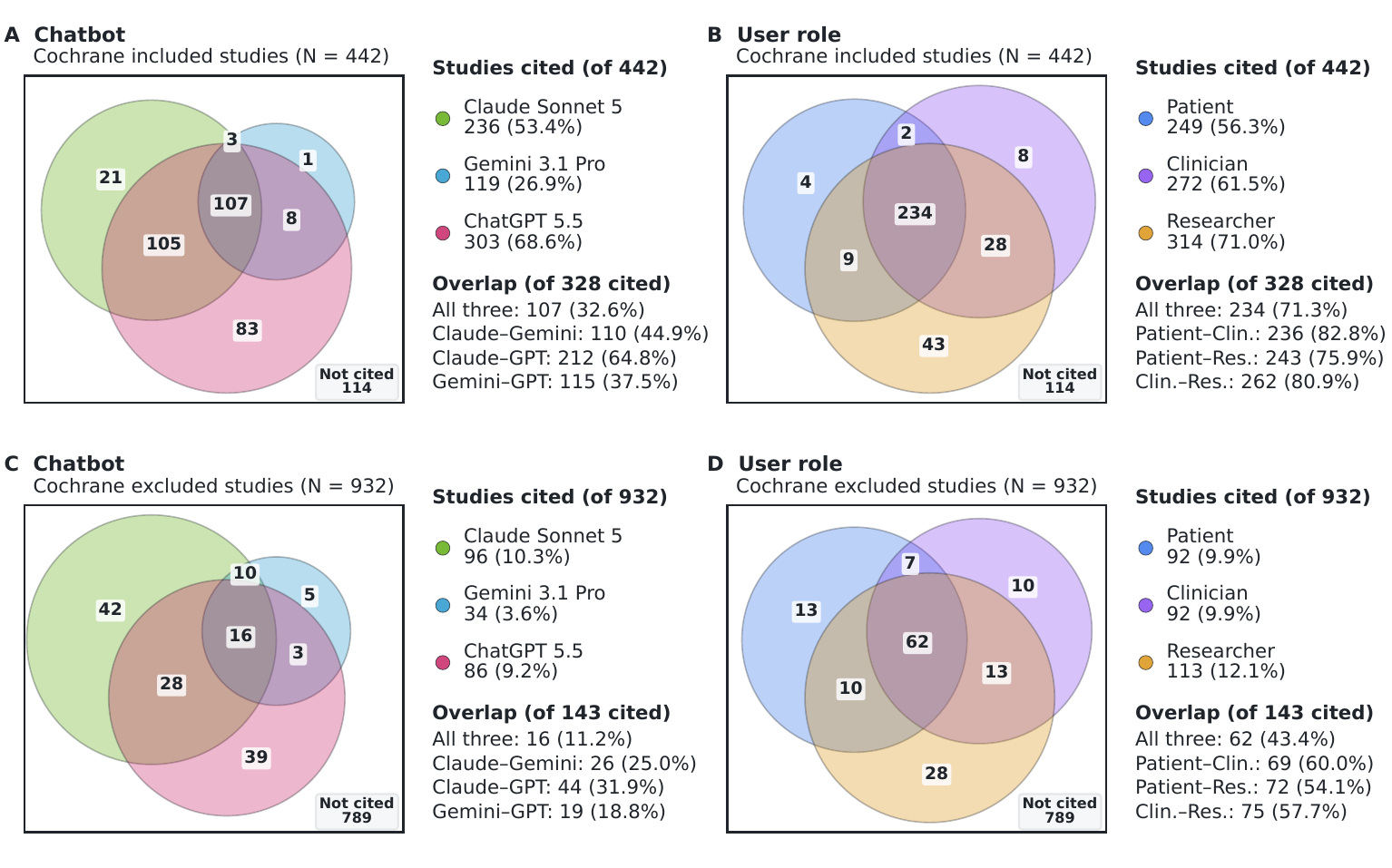}}
\caption{\textbf{Overlap in Cochrane-included and Cochrane-excluded studies cited by chatbot and user role.}
Proportional-circle Venn diagrams show citations of 442 included and 932 excluded study labels across 20 Cochrane reviews. Circle areas are proportional within each panel to the number of studies cited by each chatbot or role. Region labels report the number of studies in each mutually exclusive shared or unique citation pattern. The box enclosing each panel's circles denotes the full benchmark of included or excluded studies, and the boxed annotation at bottom right gives the number of studies not cited by any group. At right, the ``Studies cited'' list reports the same per-group counts and their percentage of the full benchmark, and the ``Overlap'' list reports the three-way and pairwise Jaccard indices (intersection over union) among the three groups.}
\label{fig:retrieval-overlap}
\end{figure}

We examined which studies were cited across multiple chatbots or user-role conditions and which were cited exclusively within a single group. Cochrane-included and Cochrane-excluded studies were analyzed separately (\textbf{Figure~\ref{fig:retrieval-overlap}}). Of the 442 Cochrane-included studies, 328 (74.2\%) were cited in at least one response. When responses were grouped by chatbot and pooled across user roles, Claude cited 236 studies (53.4\%), Gemini cited 119 (26.9\%), and ChatGPT cited 303 (68.6\%). Among the 328 cited studies, 107 (32.6\%) were cited by all three chatbots. Eighty-three studies were cited exclusively by ChatGPT, 21 exclusively by Claude, and one exclusively by Gemini; an additional 105 were cited by both Claude and ChatGPT but not by Gemini. When responses were grouped by user role and pooled across chatbots, 249 studies (56.3\%) were cited under the patient role, 272 (61.5\%) under the clinician role, and 314 (71.0\%) under the researcher role. Of the 328 cited studies, 234 (71.3\%) were cited under all three roles, whereas 4, 8, and 43 were cited exclusively under the patient, clinician, and researcher roles, respectively. Thus, the included-study sets overlapped more strongly across user roles than across chatbots.

Of the 932 Cochrane-excluded studies, 143 (15.3\%) were cited in at least one response, and only 16 of the 143 cited excluded studies (11.2\%) were cited by all three chatbots. Across user roles, 62 of the 143 cited excluded studies (43.4\%) were cited under all three roles. Thus, the excluded-study sets also overlapped more strongly across user roles than across chatbots.

\subsection{Study characteristics associated with recall}

Using the studies included in 20 Cochrane reviews as the reference standard, we compared recalled and unrecalled studies after pooling across all chatbots, user roles, and repetitions. A study was classified as recalled if it appeared in at least one response, and each Cochrane study label was counted once per review. Of 442 studies, 328 (74.2\%) were recalled and 114 (25.8\%) were not. Recalled studies were more recent (median publication year, 2014 vs.\ 2009; $p=0.00058$), larger (median analyzed sample size, 195 vs.\ 97; $p<0.001$), and more frequently cited (median citations per year, 5.50 vs.\ 2.33; $p<0.001$). Open-access status did not differ significantly between recalled and unrecalled studies (56\% vs.\ 45\%; $p=0.077$).

A multivariable logistic regression included publication year, log-transformed sample size, log-transformed citations per year, and open-access status (\textbf{Table \ref{tab:recall-logistic-regression}}). Of 442 Cochrane-included studies, 330 (74.7\%) had complete predictor data; missing values mainly reflected unavailable unresolved PMID-linked citation and open-access data from Semantic Scholar. Among the 330 studies, only sample size was significantly associated with recall, with each doubling of sample size associated with 50\% higher odds of retrieval after natural-log transformation (adjusted OR, 1.50; 95\% CI, 1.24--1.81; $p<0.001$). Publication year, citations per year, and open-access status were not significantly associated with recall.

\begin{table}[ht]
\tbl{Multivariable logistic regression of study recall.}
{\small
\begin{tabular}{@{}lrrrrrrr@{}}
\toprule
Predictor
& $\hat{\beta}$
& Clust.\ SE
& Clust.\ $p$
& Naive $p$
& OR
& 95\% CI
& VIF \\
\colrule
Intercept
& $-34.9314$
& $35.1336$
& $0.320$
& $0.194$
& ---
& ---
& --- \\
Publication year
& $0.0164$
& $0.0176$
& $0.350$
& $0.219$
& $1.017$
& $(0.98,\ 1.05)$
& $1.30$ \\
$\log(\mathrm{sample\ size})$
& $0.5880$
& $0.1390$
& $<0.001^{***}$
& $<0.001$
& $1.800$
& $(1.37,\ 2.36)$
& $1.19$ \\
$\log(\mathrm{citations/year}+0.01)$
& $0.2993$
& $0.1574$
& $0.057^{\dagger}$
& $0.010$
& $1.349$
& $(0.99,\ 1.84)$
& $1.54$ \\
Open access
& $-0.1043$
& $0.2598$
& $0.688$
& $0.740$
& $0.901$
& $(0.54,\ 1.50)$
& $1.20$ \\
\botrule
\end{tabular}}
\tabnote{\textit{Note:} The model included 330 of the 442 Cochrane-included studies with complete data for all predictors. Standard errors and primary $p$-values were clustered by review (20 review clusters); naive $p$-values ignore clustering. OR = odds ratio; CI = confidence interval; VIF = variance inflation factor. Odds ratios and confidence intervals are not reported for the intercept. McFadden pseudo-$R^2=0.1340$; log-likelihood $=-146.64$; likelihood-ratio test against the null model, $p<0.001$. $^{***}p<0.001$; $^{\dagger}p<0.10$. The tabulated OR corresponds to a one-unit increase in ln(N); the OR per doubling of N is 1.50.}
\label{tab:recall-logistic-regression}
\end{table}

\subsection{Characterizing Cochrane-excluded studies cited by chatbots}

Given that the mean excluded-set retrieval rate was $5.0\% \pm 9.4\%$ per response, we characterized the retrieved Cochrane-excluded studies by classifying the 143 studies into 11 categories using the exclusion reasons reported by Cochrane review authors (\textbf{Table~\ref{tab:exclusion_categories}}). All 143 studies were assigned to at least one category; 29 studies (20.3\%) had multiple exclusion reasons, so the categories were not mutually exclusive. The most common reasons for exclusion concerned study design, intervention, comparator, and population. This pattern is consistent with the structure of Cochrane eligibility criteria, which are typically defined according to the population, intervention, comparator, and eligible study designs specified for each review \cite{mckenzie2019defining}. Because our prompts were broader than the reviews’ full eligibility criteria, some studies could be relevant to the chatbot questions yet ineligible for the more narrowly defined reviews \cite{edinger2013large}.

\begin{table}[ht]
\tbl{Exclusion categories among 143 Cochrane-excluded studies.}
{\small
\begin{tabular}{@{}llp{0.55\textwidth}@{}}
\toprule
Category & n/143 (\%) & Definition\\
\colrule
Study design & 37 (25.9) & Nonrandomized, observational, single-arm, pseudo-randomized, post-hoc, or otherwise ineligible designs\\
Intervention & 34 (23.8) & Ineligible intervention identity, composition, route, dose, timing, or co-intervention\\
Comparator & 28 (20.3) & Absent or ineligible control, or an ineligible comparison between intervention variants\\
Population & 25 (17.5) & Ineligible participants or mixed populations without separable results for the eligible subgroup\\
Insufficient or unavailable data & 15 (10.5) & Eligibility or eligible effects could not be established, separated, or obtained\\
Timing or follow-up & 11 (7.7) & Ineligible treatment timing, study duration, or follow-up duration\\
Publication or trial status & 8 (5.6) & Protocols, registrations without eligible results, unpublished, withdrawn, or retracted trials, or duplicate or secondary reports\\
Setting & 6 (4.2) & Ineligible care or recruitment setting\\
Unspecified eligibility & 5 (3.5) & Inclusion criteria were not met, but no specific criterion was identified\\
Outcome & 2 (1.4) & Required outcomes were absent or ineligible\\
Unit of analysis & 2 (1.4) & Ineligible allocation or analysis unit, or non-independent observations\\
\botrule
\end{tabular}}
\label{tab:exclusion_categories}
\end{table}

\subsection{Hallucination, fabrication and metadata errors}

Hallucination is widely recognized as a critical issue in LLM-generated citations, but the term serves as an umbrella for several distinct phenomena \cite{walters2023fabrication,dassen2026factummechanisticdetectioncitation,bhattacharyya2023high}: (1) metadata errors (e.g., incorrect lead author, year, journal, or page range) that can nonetheless be resolved to a real publication; (2) references that cannot be resolved to any real publication; and (3) misattribution of study content to an otherwise correctly identified source. Here, we report the first two categories.

Across the 20 reviews, 232 of 7{,}676 matched response--study citation rows (3.0\%) contained metadata errors. Error rates varied substantially by model: 5.9\% for Claude (175/2{,}956), 1.5\% for ChatGPT (55/3{,}792), and 0.2\% for Gemini (2/928). Differences by user role were smaller, with error rates of 2.7\% for patients, 3.1\% for clinicians, and 3.2\% for researchers. In addition to these metadata errors, we identified 23 citations that could not be resolved to any real publication; all occurred in Claude responses.

\subsection{Secondary-evidence use and instruction adherence}

We inspected the chatbots' final responses for evidence of adherence to our instruction not to rely on secondary evidence, particularly Cochrane reviews. We screened all 720 recorded responses (240 per chatbot) and coded each response along two dimensions: (1) \textit{review identity}, indicating whether it referred to the corresponding 2026 Cochrane review, an earlier or edition-unspecified Cochrane review, or no detected Cochrane review; and (2) \textit{review role}, indicating whether the response explicitly stated that a review had been used to identify or verify primary studies or merely mentioned a review without indicating such a role. See Supplementary \textbf{Table S1} for representative examples.

Cochrane reviews appeared in 66 Claude responses (27.5\%). Of these, 17 responses (7.1\% of all Claude responses) explicitly stated that a review had been used to identify or verify primary studies, and 11 (4.6\%) referred to the corresponding 2026 Cochrane review. Only one response fell into both categories, referring to the corresponding 2026 Cochrane review while also stating that a review had been used to identify or verify primary studies. We detected no references to Cochrane reviews in the saved Gemini or ChatGPT responses. Because we did not retain complete reasoning traces or external search histories, these results reflect only what was disclosed in the final responses. A chatbot may therefore have consulted a Cochrane review without mentioning it in its final answer.

\section{Discussion}

\subsection{Study findings}

Our findings show that chatbot retrieval was incomplete and selective. A single response retrieved an average of 39.2\% of Cochrane-included studies, whereas pooling all responses identified 74.2\% at least once. Recall varied more by chatbot than by user role. Larger sample size was the only significant predictor of retrieval after accounting for publication year, citation rate, and open-access status. Fully fabricated references were relatively rare and occurred only in Claude responses, whereas citation metadata errors persisted across models and were more frequent for Claude than for ChatGPT and Gemini. Instruction adherence also varied by chatbot: Claude sometimes referenced Cochrane reviews despite instructions not to rely on secondary evidence, whereas no such references were detected in the saved ChatGPT or Gemini responses.

\subsection{Recall of primary studies}

Recall in this study was generally higher than that reported in several earlier evaluations of general-purpose chatbots: mean per-response recall was 39.2\% overall and 63.1\% for ChatGPT, compared with 0\%--13.7\% in Chelli et al.\ \cite{chelli2024hallucination}, only one and two of 24 benchmark trials in Gwon et al.\ \cite{gwon2024scientificsearches}, and 35.7\% for the best-performing model in Somer et al\ \cite{somer2026studyidentification}. This difference may reflect the use of newer LLMs with stronger reasoning and agentic capabilities, an explicit prompt requesting a list of primary studies, and a user role closely matched to the medical question.

We observed differences in recall across different chatbot models, and this appears to be largely driven by citation volume. ChatGPT cited a mean of 15.80 studies per response, approximately four times that of Gemini's 3.87. The official release materials for all three models (GPT-5.5, Claude Sonnet 5, and Gemini 3.1 Pro) emphasize advanced reasoning and agentic capabilities, and their performance on Humanity’s Last Exam was broadly comparable, especially under no-tool evaluation conditions. Despite these similarities in general reasoning performance, the models exhibited substantially different study-retrieval behaviors when answering medical questions. These discrepancies raise the question of whether they stem from differences in training data, retrieval architecture, or reasoning architecture, though the proprietary nature of these models makes this difficult to determine.

\subsection{Factors affecting retrieval}

Previous studies have reported that LLM-generated references tend to favor highly cited publications \cite{tang2025literaturereview,algaba2025citationbias}. In our study, however, after accounting for publication year, citations per year, and open-access status, trial size emerged as the clearest predictor of retrieval, with larger trials more likely to be retrieved. Citation rate was also positively associated with retrieval, but this association did not reach statistical significance, as the clustered confidence interval included the null ($p=0.057$). Several mechanisms may explain the preferential retrieval of larger studies. Search or retrieval-augmented generation pipelines may rank prominent ``landmark'' trials more highly, while larger trials may also be more strongly represented in model training data or other web-accessible sources. Chatbots may further select larger trials because sample size is interpreted as a signal of stronger or more reliable evidence. Alternatively, sample size may serve as a proxy for unmeasured characteristics such as journal visibility, trial registration, and broader scientific prominence. Because the search, retrieval, and reranking processes of these consumer chatbot systems are proprietary and opaque, the current study cannot distinguish those possible mechanisms.

\subsection{Implications}

Our study has several important implications. First, patients and clinicians should recognize that a chatbot response with citations reflects a model-specific and role-conditioned subset of the literature that may vary across repeated queries, rather than a comprehensive evidence summary. Second, systematic reviewers and evidence-synthesis researchers may find that general-purpose chatbots identify some relevant studies, but these tools should not replace reproducible database searches, formal eligibility screening, or citation verification \cite{clark2025generative,chen2025can}. Third, AI researchers and developers working to automate evidence synthesis should make efforts to improve how systems interpret and apply population, intervention, comparator, outcome, and study-design (PICO-S) criteria. Accurate application of these criteria is essential for distinguishing the retrieval of high-quality evidence from the comprehensive retrieval of all relevant evidence \cite{li2025enhancing,vallamchetla2025faster}.

\subsection{Limitations}

Our study has several limitations. First, querying the models with questions from only 20 recent reviews limits the findings' breadth and generalizability, making this an exploratory test set rather than a comprehensive benchmark. Also, although Cochrane reviews provide a rigorous reference standard, their included-study sets are not immutable and may vary with review-specific judgments \cite{hamilton2026evaluation}. Second, the study may not fully capture real-world patterns of use. Prompts were kept nearly identical across user roles to isolate the effect of this variable, but patients may express their evidence needs in more varied, less structured ways in practice \cite{roberts2016interactive,zeng2002characteristics}. Models were also instructed to avoid secondary evidence to facilitate comparison with the Cochrane reference standard, but in real life, users are unlikely to add such constraints. Third, although we selected Cochrane reviews published after each model's knowledge cutoff to minimize potential training data exposure, the models may still have been exposed to earlier versions or other secondary sources. Moreover, despite being explicitly instructed to avoid using secondary evidence, the models differed in their adherence to this constraint \cite{jiang-etal-2024-followbench}.

\subsection{Future work}

We propose several aspects of future work. First, future work should evaluate citations that fall outside the sets of included and excluded studies. Based on the present analysis, we cannot determine whether these citations represent irrelevant studies retrieved in error, newly published eligible studies, or other forms of related evidence \cite{garner2016and}. Assessing their relevance will require independent adjudication and, in some cases, clinical expertise. Second, because our analysis focused on study retrieval, future research should also examine whether chatbot-generated conclusions accurately reflect the underlying evidence \cite{wu2025sourcecheckup,jin2026med}, appropriately balance benefits and harms, and communicate uncertainty. Such evaluations will likely require topic-specific expertise, particularly when the evidence is heterogeneous or clinically complex. Third, our analysis of cited study characteristics focused on publication year, trial size, citation rate, and open-access status. Additional factors such as the authors' country and abstract length warrant further study. Finally, future work should directly compare chatbot-based retrieval with conventional search-engine approaches, as well as newer agentic search systems, including Deep Research in ChatGPT and Gemini and domain-specific agentic RAG systems, to determine whether they can retrieve additional studies included in the reference Cochrane reviews that were missed by the chatbot systems evaluated in the current work \cite{wang2026empowering,wang2026deepermedadvancingdeepevidencebased,xi-etal-2026-survey,deng-etal-2026-data}.

\section{Acknowledgements}

This research was supported by the Intramural Research Program of the National Institutes of Health (NIH). The contributions of the NIH author(s) are considered works of the US Government. Q.J. was also supported by the NIH Pathway to Independence Award K99LM014903. The findings and conclusions presented in this paper are those of the author(s) and do not necessarily reflect the views of the NIH or the US Department of Health and Human Services.

\bibliographystyle{ws-procs11x85}
\bibliography{retrieval_bias_literature}

\end{document}